\documentclass{sig-alternate}
\usepackage{graphicx}
\usepackage{colortbl}
\usepackage{subfigure}
\usepackage{multirow}
\usepackage{url}

\begin{document}
\conferenceinfo{JCDL'10,} {June 21--25, 2010, Gold Coast, Queensland, Australia.} 
\CopyrightYear{2010}
\crdata{978-1-4503-0085-8/10/06}
\clubpenalty=10000
\widowpenalty = 10000

\title{Evaluating Methods to Rediscover Missing Web Pages from the Web Infrastructure}
\numberofauthors{2}
\author{
% 1st. author
\alignauthor
Martin Klein\\
       \affaddr{Department of Computer Science}\\
       \affaddr{Old Dominion University}\\
       \affaddr{Norfolk, VA,  23529}\\
       \email{mklein@cs.odu.edu}
% 2nd. author
\alignauthor
Michael L. Nelson\\
       \affaddr{Department of Computer Science}\\
       \affaddr{Old Dominion University}\\
       \affaddr{Norfolk, VA,  23529}\\
       \email{mln@cs.odu.edu}
}
\maketitle
\begin{abstract}
Missing web pages (pages that return the $404$ ``Page Not Found'' error) are part of the browsing experience.
The manual use of search engines to rediscover missing pages can be frustrating and unsuccessful.
We compare four automated methods for rediscovering web pages. We extract the page's title, generate the page's lexical signature (LS),
obtain the page's tags from the bookmarking website \texttt{delicious.com} and generate a LS from the page's link neighborhood.
We use the output of all methods to query Internet search engines and analyze their retrieval performance.
Our results show that both LSs and titles perform fairly well with over $60\%$ URIs returned top ranked from Yahoo!. However, the
combination of methods improves the retrieval performance. Considering the complexity of the LS generation, querying the title first and
in case of insufficient results querying the LSs second is the preferable setup. This combination accounts for more than $75\%$ top ranked
URIs.
\end{abstract}

\category{H.3.3}{Information Storage and Retrieval}Information Search and Retrieval
\terms{Measurement, Performance, Design, Algorithms}
\keywords{Web Page Discovery, Digital Preservation, Search Engines}

\section{Introduction} \label{sec:intro}
Inaccessible web pages and ``404 Page Not Found'' responses are part of the web browsing experience.
Despite guidance for how to create ``Cool URIs'' that do not change \cite{berners-lee:cool} there are many reasons why URIs
or even entire websites break \cite{marshall:archiving_strategies}.
However, we claim that information on the web is rarely completely lost, it is just missing.
In whole or in part, content is often just moving from one URI to another. It is our intuition that major search engines
like Google, Yahoo! and MSN Live (our experiments were conducted before Microsoft introduced Bing), as members of what we call
the Web Infrastructure (WI), likely have crawled the content and
possibly even stored a copy in their cache. Therefore the content is not lost, it ``just'' needs to be rediscovered.
The WI, explored in detail in \cite{jatowt:browser,mccown:thesis,nelson:web-infrastructure}, also includes (besides
search engines) non-profit archives such as the Internet Archive (IA) or the European Archive as well as large-scale academic
digital data preservation projects e.g., CiteSeer and NSDL.

\begin{figure}[ht]
\begin{center}
 \subfigure[Original URI, new (unrelated) Content]{\label{fig:ht06_1}\includegraphics[scale=0.2]{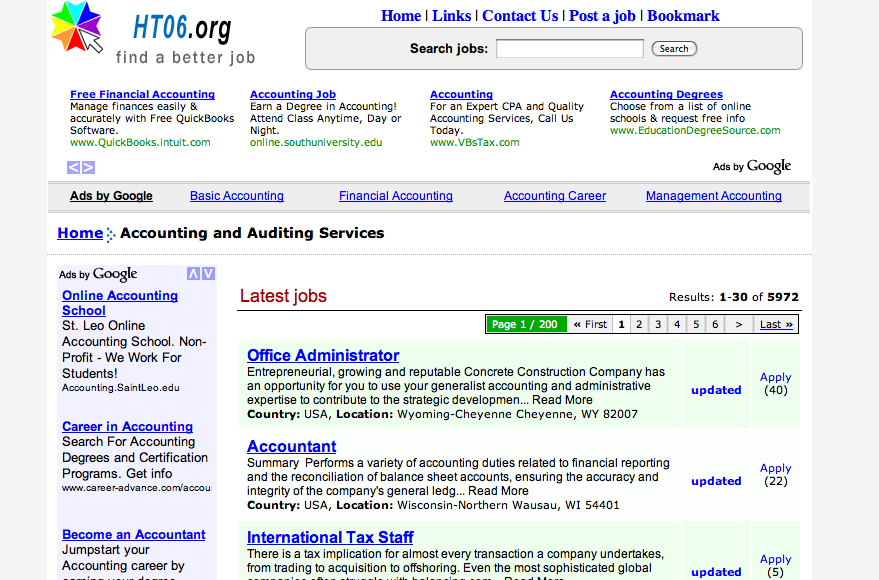}}
 \vspace{.1in}
 \subfigure[Original Content, new URI]{\label{fig:ht06_2}\includegraphics[scale=0.2]{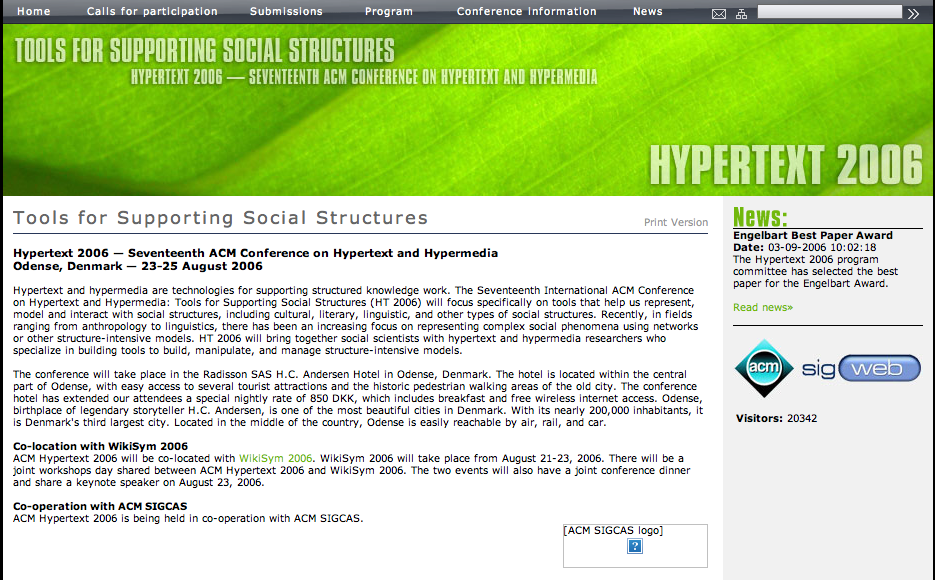}}
 \caption{The Content of the Website for the Conference Hypertext $2006$ has Moved over Time}
 \label{fig:ht06}
\end{center}
\end{figure}

It is commonplace for content to ``move'' to different URIs over time.
Figure \ref{fig:ht06} shows two snapshots as an example of a web page whose content has moved within one year after its creation.
Figure \ref{fig:ht06_1} shows the content of the original URI of the Hypertext $2006$ conference\footnote{\texttt{http://www.ht06.org/}}
as displayed in 12/2009.
The original URI clearly does not hold conference related content anymore. Our suspicion is that the website administrators did not renew
the domain registration and therefore enabling someone else to take over. However, the content is not lost.
It is now available at a new URI\footnote{\texttt{http://hypertext.expositus.com/}} as shown in Figure \ref{fig:ht06_2}.

In this paper we investigate the retrieval performance of four methods that can be automated and together with the WI
used to discover missing web pages. These methods are:
\begin{enumerate}
\item lexical signatures (LSs) -- typically the $5$-$7$ most significant keywords extracted from a cached copy of the missing page that capture
its ``aboutness''
\item the title of the page -- the two underlying assumptions here are:
 \begin{itemize}
 \item web pages have descriptive titles
 \item titles only change infrequently over time
 \end{itemize}
\item social bookmarking tags -- terms suggested by Internet users on \texttt{delicious.com} when the page was bookmarked
\item link neighborhood LSs (LNLS) -- a LS generated from the pages that link to the missing page (inlinks) and not from a cached copy of the missing page.
\end{enumerate}

\begin{figure}[ht]
 \center
 \includegraphics[scale=0.45]{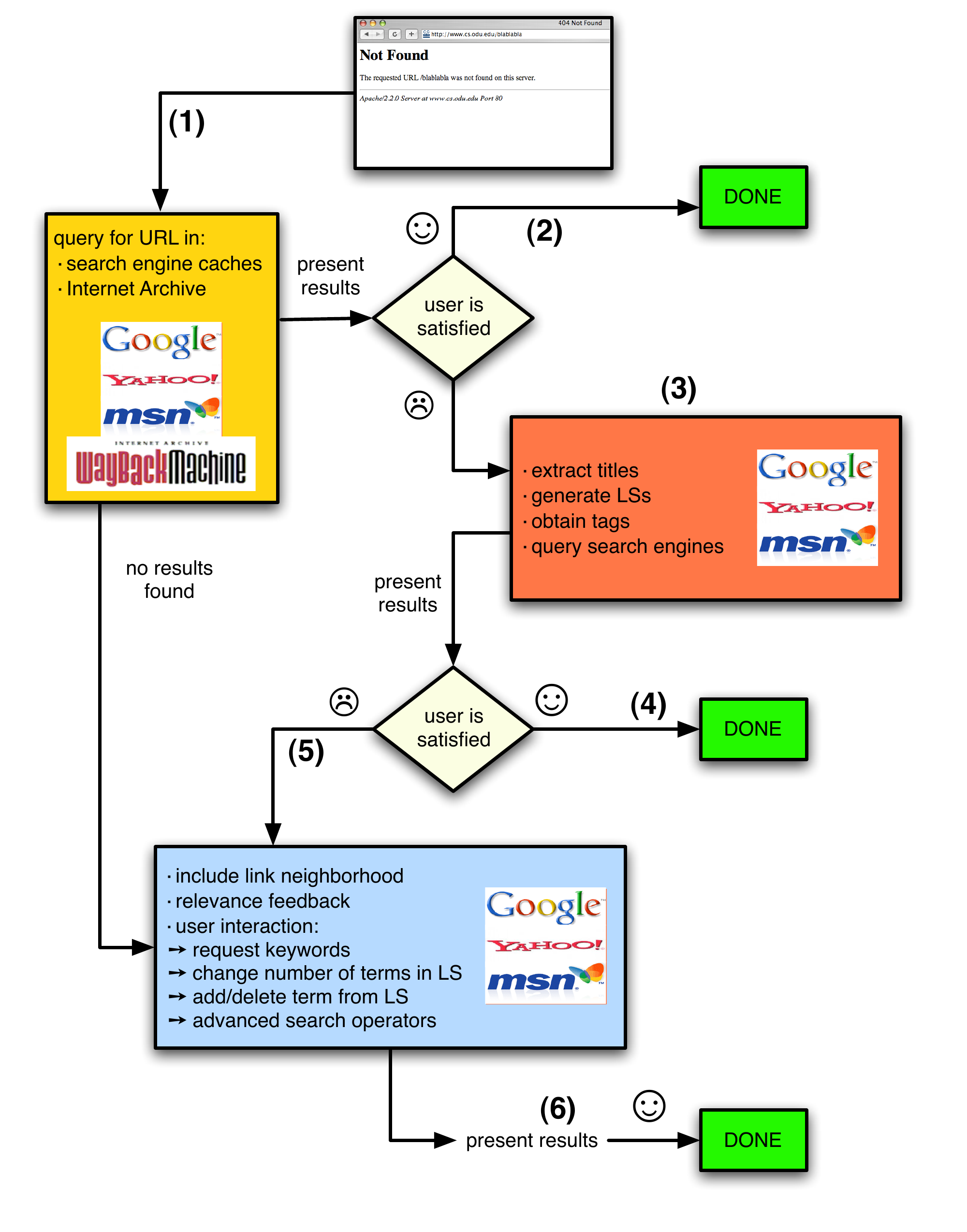}
 \caption{Process to Rediscover Missing Web Pages}
 %\caption{Flow Diagram to Rediscover Missing Web Pages}
 \label{fig:flow_diagram}
\end{figure}

Figure \ref{fig:flow_diagram} displays the scenario how the four methods of interest can automatically be applied for the discovery of a missing page.
The occurrence of an $404$ error is displayed in the first step. % of Figure \ref{fig:flow_diagram}.
Search engine caches and the IA will consequently be queried with the URI requested by the user. In case older copies of the page
are available they can be offered to the user. If the user's information need is satisfied, nothing further needs to be done (step ($2$)).
If this is not the case we need to proceed to step ($3$) where we extract titles, try to obtain tags about the URI and generate LSs from the
obtained copies. The obtained terms are then queried against live search engines. %as depicted in step ($4$) of Figure \ref{fig:flow_diagram}.
The returned results are again offered to the user and in case the outcome is not satisfying more sophisticated and complex methods
need to be applied (step ($5$)).
Search engines can be queried to discover pages linking to the missing page. The assumption is that the aggregate of those pages 
is likely to be ``about'' the same topic. From this link neighborhood a LS can be generated. At this point the approach is the same as the LS
method, with the exception that the LS has been generated from a link neighborhood and not a cached copy of the page itself.
The important point of this scenario is that it works while the user is browsing and therefore has to provide results in real time.
Queries against search engines can be automated through APIs but the generation of LSs needs to be automated too. 

As an example let us look at how the methods would be applied to the web page \texttt{www.nicnichols.com}.
The page is about a photographer named Nic Nichols.
Table \ref{tab:data_example} displays all data we obtained about the page using the four methods.
%, the LS, title, tags and
%link neighborhood based LS (LNLS).
%
\begin{table}
  \begin{tabular}{|l||l|} \hline
        \multirow{0}{*}{\textbf{LS}}&NICNICHOLS NICHOLS NIC STUFF \\
	&SHOOT COMMAND PENITENTIARY \\ \hline
        \multirow{0}{*}{\textbf{Title}}&NICNICHOLS.COM : DOCUMENTARY \\
	&TOY CAMERA PHOTOGRAPHY OF \\
	&NIC NICHOLS : HOLGA, LOMO \\
	&AND OTHER LO-FI CAMERAS! \\ \hline
        \multirow{0}{*}{\textbf{Tags}}&PHOTOGRAPHY BLOG PHOTOGRAPHER \\
	& PORTIFOLIO PORTFOLIO INSPIRATION \\
	&PHOTOGRAPHERS \\ \hline
        \multirow{0}{*}{\textbf{LNLS}}&NICNICHOLS PHOTO SPACER \\
	&VIEW PHIREBRUSH SUBMISSION \\
	&BOONIKA \\ \hline
  \end{tabular}
  \caption{Data Obtained from \texttt{www.nicnichols.com}}
 \label{tab:data_example}
\end{table}
%The overlap and difference between the data is obvious, the question now is, what data performs best for discovering the page, in case it has
%gone missing.
The question is now if \texttt{nicnichols.com} went missing (returned HTTP $404$ response code), which of the four methods will produce the best search engine
query to rediscover Nic Nichol's website if it moved to a new URI?
To further illustrate the difference between titles and LSs we compare their retrieval performance with the following three examples.
From the URI \url{smiledesigners.org} we derive a LS and a title ($T$):
\begin{itemize}
\item LS: \emph{``Dental Imagined Pleasant Boost Talent Proud Ways''}% and
\item T: \emph{``Home''}.
\end{itemize}
When queried against Google the LS returns the URI top ranked but since the title is rather arbitrary it does not return the URI
within the top $100$ results.
From the the URI \url{www.redcrossla.org} we get
\begin{itemize}
\item LS: \emph{``Marek Halloween Ready Images Schwarzenegger Govenor Villaraigosa''}% and 
\item T: \emph{``American Red Cross of Greater Los Angeles''}.
\end{itemize}
The LS contains terms that were part of the page at the time of the crawl but are less descriptive. Hence the URI remains undiscovered.
The title of the page however performs much better and returns the URI top ranked.
From the last example URI \url{www.aircharter-international.com} we obtain
\begin{itemize}
\item LS: \emph{``Charter Aircraft Jet Air Evacuation Medical Medivac''}% and
\item T: \emph{``ACMI, Private Jet Charter, Private Jet Lease, Charter Flight Service: Air Charter International''}.
\end{itemize}
Both describe the page's content very well and return the URI top ranked.

The contribution of this paper is the performance comparison of all our methods and an interpretation resulting in a suggested workflow
on how to set up the investigated methods to achieve a highest possible rate in discovering missing web resources.
\section{Related Work} \label{sec:relwork}
%
%\subsection{Missing Web Pages}
\subsection{Missing Web Resources}
Missing web pages are a pervasive part of the web experience.
The lack of link integrity on the web has been addressed by numerous researchers
\cite{ashman:document_addressing,ashman:missing404,davis:ht_link_integrity,davis:referential_integrity}.
In $1997$ Brewster Kahle published an article focused on preservation of Internet resources claiming that the expected lifetime of a web page is $44$
days \cite{kahle:preserving}.
A different study of web page availability performed by Koehler \cite{koehler:web-page-change} shows the random test collection of URIs
eventually reached a ``steady state'' after approximately $67$\% of the URIs were lost over a $4$-year period.
Koehler estimated the half-life of a random web page is approximately two years.
Lawrence et al. \cite{lawrence:persistence} found in $2000$ that between $23$ and $53\%$ of all URIs occurring in computer science
related papers authored between $1994$ and $1999$ were invalid. By conducting a multi level and partially manual search on the Internet,
they were able to reduce the number of inaccessible URIs to $3$\%. This confirms our intuition that information is rarely lost, it is just moved.
This intuition is also supported by Baeza-Yates et al. \cite{baeza-yates:genealogical_trees} who show that a significant portion of the web
is created based on already existing content.

Spinellis \cite{spinellis:decay} conducted a study investigating the accessibility of URIs occurring in papers published in
Communications of the ACM and IEEE Computer Society. He found that $28$\% of all URIs were unavailable after five years and $41\%$ after seven years.
He also found that in $60\%$ of the cases where URIs where not accessible, a $404$ error was returned. He estimated the half-life of an URI
in such a paper to be four years from the publication date.
Dellavalle et al. \cite{dellavalle:going} examined Internet references in articles published in journals with a high impact factor given by the
Institute for Scientific Information (ISI).
They found that Internet references occur frequently (in $30\%$ of all articles) and are often inaccessible within months after publication
in the highest impact (top $1\%$) scientific and medical journals. They discovered that the percentage of inactive references (references that return
an error message) increased over time from $3.8\%$ after $3$ month to $10\%$ after $15$ month up to $13\%$ after $27$ month.
The majority of inactive references they found were in the \emph{.com} domain ($46\%$) and the fewest in the \emph{.org} domain ($5\%$).
By manually browsing the IA they were able to recover information for about $50\%$ of all inactive references.

Zhuang et al. \cite{zhuang:focused_crawling} and Silva et al. \cite{Silva:missing_from_diglib} have used the web infrastructure
to obtain missing documents from digital library collections. Their notion of ``missing documents'' however is different from ours since
they focus on enhancing existing library records with related (full text) documents. They extract the title, names of authors and
publication venues from the library records and use them as search engine queries in order to obtain resources that are not held in the
digital library.
\subsection{Search Engine Queries}
The work done by Henzinger et al. \cite{henzinger:query-free} is related in the sense that they tried to determine the ``aboutness'' of news documentations.
They provide the user with web pages related to TV news broadcasts using a $2$-term summary which can be thought of as a LS.
This summary is extracted from closed captions of the broadcast and various algorithms are used to compute the scores determining the most relevant terms.
The terms are used to query a news search engine while the results must contain all of the query terms.
The authors found that $1$-term queries return results that are too vague and $3$-term queries return too often zero results.
Thus they focus on creating $2$-term queries.

He and Ounis' work on query performance prediction \cite{he:inferring_query_performance} is based on the TREC dataset. They measured retrieval performance of
queries in terms of average precision (AP) and found that the AP values depend heavily on the type of the query.
They further found that what they call \textit{simplified clarity score (SCS)} has the strongest correlation with AP for title queries (using the title of
the TREC topics).
SCS depends on the actual query length but also on global knowledge about the corpus such as document frequency and total number of tokens in the corpus.
%Since our study is focued on discovering web pages and not documents in a bounded and static dataset, the values to compute SCS could only be estimated.
%
\subsection{The Web Infrastructure for the Preservation of Web Pages}
Nelson et al. \cite{nelson:web-infrastructure} present various models for the preservation of
web pages based on the web infrastructure. They argue that conventional approaches to digital preservation such as storing digital data in archives
and applying methods of refreshing and migration are, due to the implied costs, unsuitable for web scale preservation.

McCown has done extensive research on the usability of the web infrastructure for reconstructing missing websites \cite{mccown:thesis}.
He also developed \textit{Warrick} \cite{mccown:lazyp}, a system that crawls web repositories such as search engine caches (characterized
in \cite{mccown:se-caches}) and the index of the IA to reconstruct websites. His system is targeted to individuals and small scale communities that
are not involved in large scale preservation projects and suffer the loss of websites.
\subsection{Lexical Signatures of Web Pages}
So far, little research has been done in the field of lexical signatures for web resources. Phelps and Wilensky \cite{phelps:hyperlinks} first proposed
the use of LSs for finding content that had moved from one URI to another. Their claim was ``robust hyperlinks cost just 5 words each'' and their
preliminary tests confirmed this. The LS length of $5$ terms however was chosen somewhat arbitrarily. Phelps and Wilensky proposed ``robust hyperlinks'',
an URI with a LS appended as an argument. They conjectured that if an URI would return a $404$ error, the browser would use the LS
appended to the URI and submit it to a search engine in order to find the relocated copy.

Park et al. \cite{park:ls-tois} expanded on the work of Phelps and Wilensky, studying the performance of $9$ different LS generation algorithms
(and retaining the $5$-term precedent). The performance of the algorithms depended on the intention of the search.
Algorithms weighted for term frequency (TF; ``how often does this word appear in this document?'') were better at finding related pages, but the exact page
would not always be in the top N results.
Algorithms weighted for inverse document frequency (IDF; ``in how many documents of the entire corpus does this word appear?'') were better at finding the
exact page but were susceptible to small changes in the document (e.g., when a misspelling is fixed).
\section{Experiment Setup} \label{sec:ex_setup}
We are not aware of a data corpus providing missing web pages. Therefore we need to generate a dataset of URIs taken from the live web and ``pretend''
they are missing. We know they are indexed by search engines so by querying the right terms, we will be able to retrieve them in the result set.
\subsection{Data Gathering}
As shown in \cite{henzinger:url_sampling,rusmevichientong:sampling_pages,theall:methodologies}, finding a small sample set of URIs that represent the Internet
is not trivial. 
Rather than attempt to get an unbiased sample, we randomly sampled $500$ URIs from the Open Directory Project \texttt{dmoz.org}.
We are aware of the implicit bias of this selection but for simplicity it shall be sufficient.
We dismissed all non-English language pages as well as all pages containing
less than $50$ terms (this filter was also applied in \cite{klein:ls,park:ls-tois}). Our final sample set consists of a total of $309$ URIs, $236$ in the .com,
$38$ .org, $27$ .net and $8$ in the .edu domain.
We downloaded the content of all pages and excluded all non-textual content such as HTML and JavaScript code.
\subsection{Lexical Signature Generation} \label{subsec:ls_generation}
The LS generation is commonly done following the well known and established TF-IDF term weighting concept.
TF-IDF extracts the most significant terms from textual content while also dismissing more common terms such as stop words. It is often used for
term weighting in the vector space model as described by Salton et al. \cite{salton:vsm}.
For the IDF computation, two values are mandatory: the overall number of documents in the corpus and the number of documents, the particular term occurs in.
Both values can only be estimated when the corpus is the entire web. As a common approach researchers use search engines to estimate the document frequency
of a term (\cite{harrison:just-in-time,keller:using_the_web,phelps:hyperlinks,zhu:improving_trigram}). Even though the obtained values are only
estimates (\cite{google:df_estimates})
our earlier work \cite{klein:idf} has shown that this approach actually works well compared to using a modern text corpus.

Recent research \cite{harrison:just-in-time,klein:ls,park:ls-tois,wan:wordrank} has shown that a
LS generated from the content of the potentially missing web page can be used as a query for the WI trying to rediscover the page.
A LS is generally defined as the top $n$ terms of the list of terms ranked by their TF-IDF values in decreasing order.
We have shown in \cite{klein:ls} that $5$- and $7$-term LSs perform best, depending on whether the focus is on obtaining
the best mean rank or the highest percentage of top ranked URIs.

Our first experiment investigates the differences in retrieval performance between LSs generated from three different search engines. We use the Google, Yahoo! (BOSS)
and MSN Live APIs to determine the IDF values and compute TF-IDF values of all terms. Due to the results of our earlier research we use $5$- and $7$-term LSs for each
URI and query them against the search engine the LS was generated from. A comparison of the retrieval results from cross search engines queries was not the focus of
this paper but can be found in \cite{klein:cross_se_ls}.
As an estimate for the overall number of documents in the corpus (the Internet) we use values obtained from \cite{www:size}.
The TF-IDF score of very common words are very low with a sufficiently large corpus. Therefore these terms very likely would not make it into the top $n$ from which
a LS is generated. Despite that and for keeping the queries to determine the document frequency value low we dismiss stopwords from the textual content of the web
pages before computing TF-IDF values.
We also ran experiments with stemming algorithms applied but the resulting LSs performed very poorly and hence we decided not to report the numbers here.
\subsection{Title and Tags Extraction}
Titles of web pages seem to be commonplace. To confirm this intuition we randomly sampled another set of URIs from \texttt{dmoz.org} 
(a total of $10,000$ URIs) and parsed their content for the title.
The statistics showed that the vast majority of URIs contained a title and in only $1.1\%$ of all cases no title could be discovered.
Since we already downloaded all of our URIs, extracting the title is simply done by parsing the page and extract everything between the HTML
tags \texttt{<title></title>}.

Tags used to annotate an URI have been shown to be useful for search and even possibly hold additional information that can
contribute to search but is not included in the text of the page \cite{bischoff:tags_for_search}.
We therefore use the API provided by the bookmarking site \texttt{delicious.com} to obtain all available tags for our set of URIs.
These annotations are given by Internet users and hence may vary greatly in quality. The \texttt{delicious.com} API provides up to ten
terms per URI and they are ordered by frequency of use, supposedly indicating the importance of the term.
\subsection{Page Neighborhood Content} \label{subsec:page_nh}
As shown in \cite{sugiyama:refinement-of-tfidf} the content of neighboring pages can help retrieve the centroid page. 
Their research is based on the idea that the content of a centroid web page is often related to the content of its neighboring pages.
This assumption has been proven by Davidson in \cite{davidson:topical_locality} and Dean and Henzinger in \cite{dean:finding_related_pages}.

We download up to $50$ inlink pages (pages which have a reference to our, the centroid, page) that the Yahoo! API provides for each of our $309$ URIs. 
We generate a bucket of words from the neighborhood of each URI and apply the same procedure as in \ref{subsec:ls_generation} to generate one LS per page neighborhood.
More than $425,000$ queries were necessary to determine document frequency values of the entire neighborhood.
Since the Google API is restricted to $1000$ queries per day and the MSN Live API was in our experiments not sufficiently reliable for such a query volume 
we only use the Yahoo! API for this experiment.
\section{Experiment Results} \label{sec:expres}
As laid out in Section \ref{sec:ex_setup} we use $5$- and $7$-term LSs, titles and tags as queries to three different search engines.
We parse the top $100$ returned results for the source URI and distinguish between 4 scenarios:
\begin{enumerate}
\item the URI returns top ranked
\item the URI returns in the top $10$ but not top ranked
\item the URI returns in the top $100$ but not top ranked and not in the top $10$
\item the URI is considered not to return.
%\item the URI does not return, neither top ranked, nor in the top $10$ or top $100$.
\end{enumerate}
In the last scenario we consider the URI as undiscovered since the majority of the search engine users do not browse through the result set past the top
$100$ results. We are aware of the possibility that we are somewhat discriminating URIs that may be returned just beyond rank $100$ but we apply that
threshold for simplicity.
With these scenarios we evaluate our results as success at $1$, $10$ and $100$. Success is defined as a binary value, as the target either occurs in
the subset (top result, top $10$, top $100$) of the entire result set or it does not.
\subsection{Comparing the LS Performance}
\begin{figure}[ht]
 \center
 \includegraphics[scale=0.4]{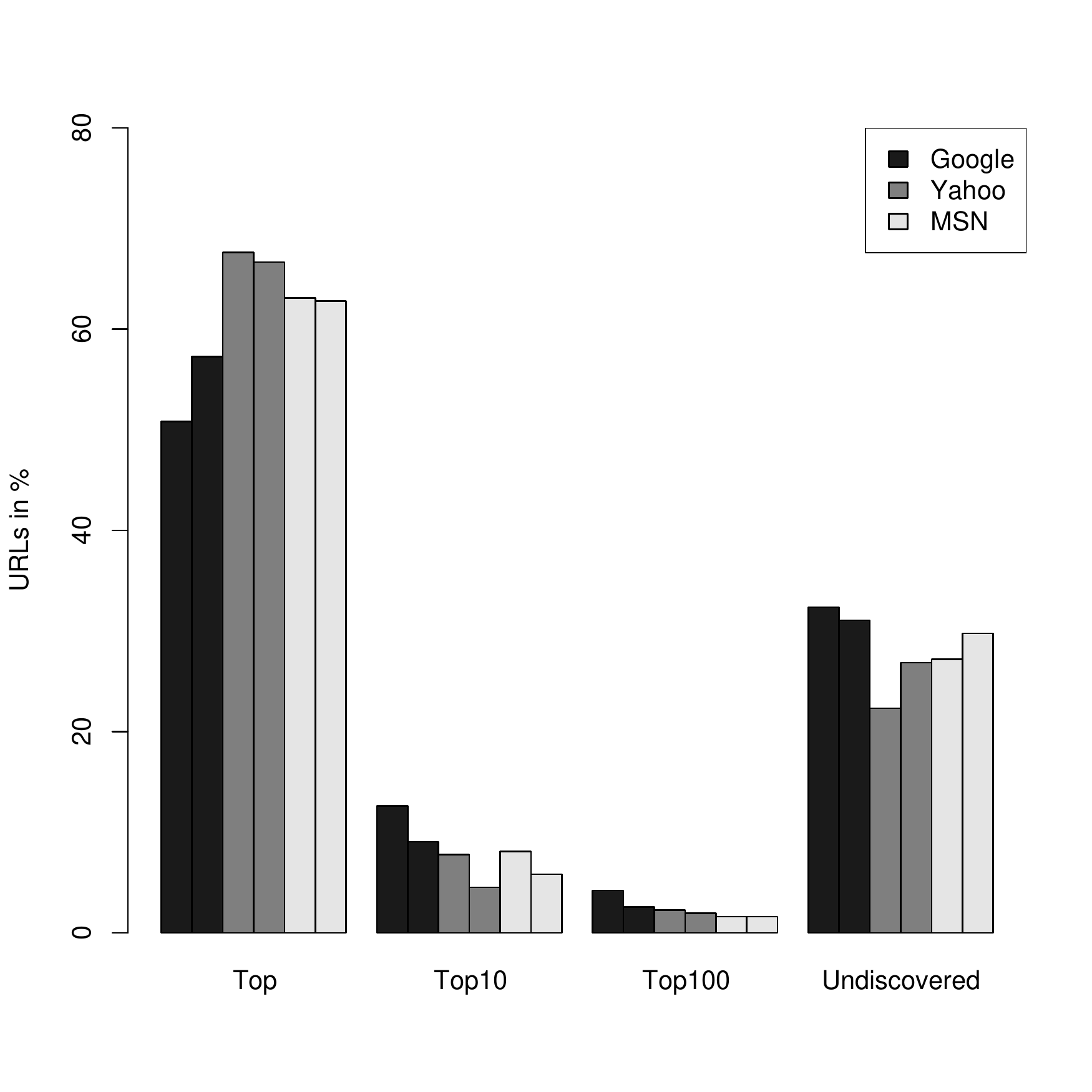}
 \caption{$5$- and $7$-Term LS Retrieval Performance}
 \label{fig:ls_retrieval}
\end{figure}

Figure \ref{fig:ls_retrieval} shows the percentage of URIs retrieved top ranked, ranked in the top $10$ and top $100$ as well as the percentage of URIs that
remained undiscovered when using $5$- and $7$-term LSs. 
For each of the four scenarios we show three tuples distinguished by color, indicating the search engine the LS was generated from and queried against.
The left bar of each tuple represents the results for $5$- and the right for $7$-term LSs.
We can observe an almost binary pattern meaning the majority of the URIs are either returned ranked between one and ten or are undiscovered.
If we for example consider $5$-term LSs fed into Yahoo! we retrieve $67.6\%$ of all URIs top ranked, $7.7\%$ ranked in the top $10$ (but not top) and 
$22\%$ remain undiscovered.
Hence the binary pattern: we see more than $75\%$ of all URIs ranked between one and ten and vast majority of the remaining
quarter of URIs was not discovered.
Yahoo! returns the most URIs and leaves the least undiscovered. MSN Live, using $5$-term LSs, returns more than $63\%$ of the URIs as the top result and hence
performs better than Google which barely returns $51\%$.
Google returns more than $6\%$ more top ranked results with $7$-term LSs compared to when $5$-term LSs were used.
Google also had more URIs ranked in the top $10$ and top $100$ with $5$-term LSs. These two observations confirm our findings in \cite{klein:ls}.
\subsection{Performance of Titles}
\begin{figure}[ht]
 \center
 \includegraphics[scale=0.4]{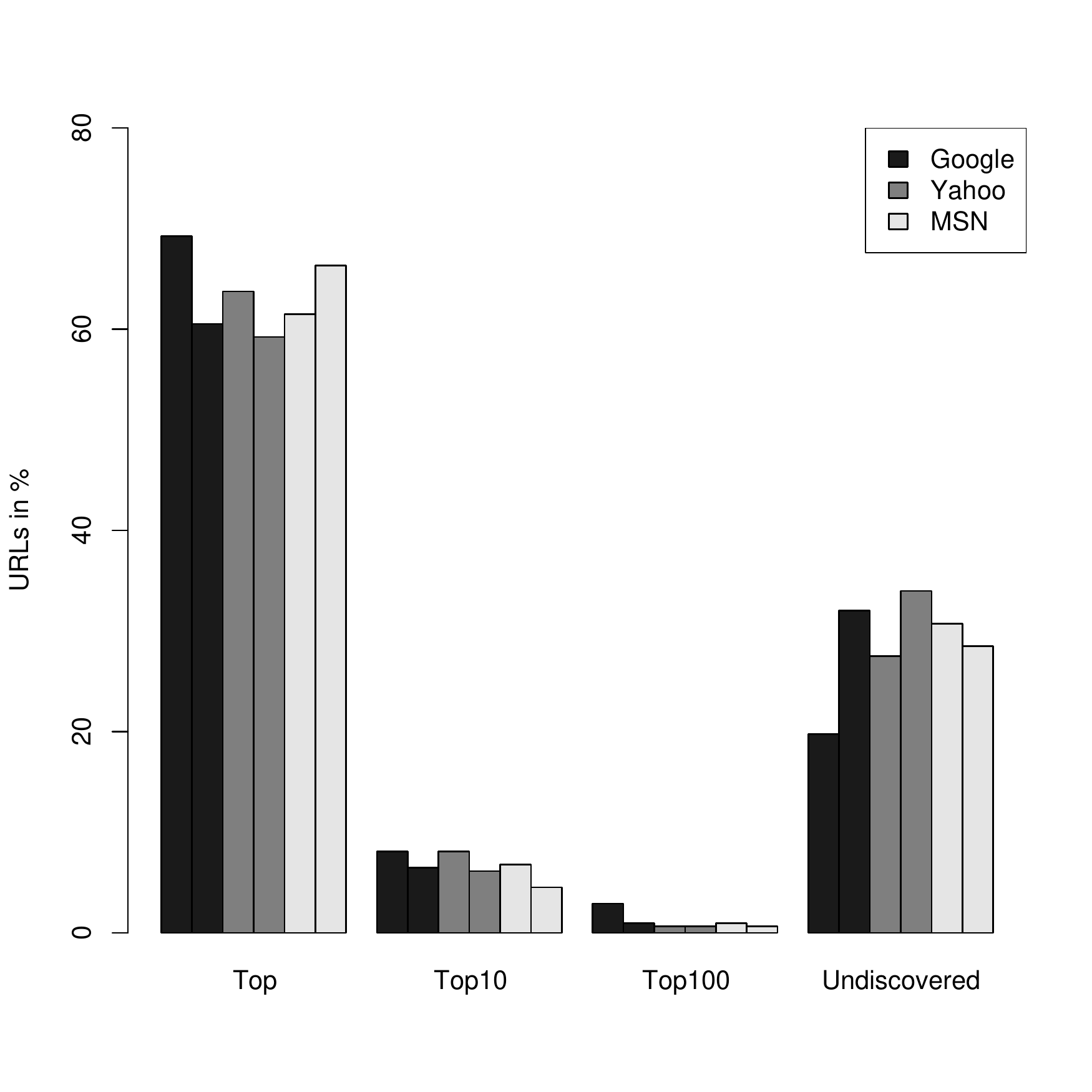}
 \caption{Non-Quoted and Quoted Title Retrieval Performance}
 \label{fig:title_retrieval}
\end{figure}

The bars displayed in Figure \ref{fig:title_retrieval} show the percentages of retrieved URIs when querying the title of the pages. We queried the title once without
quotes and once quoted, forcing the search engines to handle all terms of the query as one string. The left bar of each tuple (again distinguished by color) shows
the results for the non-quoted titles.
To our surprise both Google and Yahoo! return fewer URIs when using quoted titles. Google in particular returns $14\%$ more top
ranked URIs and $38\%$ less undiscovered URIs for the non-quoted titles compared to the quoted titles.
Only MSN Live shows a different behavior with more top ranked results (almost $8\%$ more) for the quoted and more undiscovered URIs (more than $7\%$) using the
non-quoted titles.
We can see however that titles are a very well performing alternative to LSs. The top value for LSs was obtained from Yahoo! ($5$-term) with $67.6\%$ top ranked URIs
returned and for titles with Google (non-quoted) which returned $69.3\%$ URIs top ranked.
\subsection{Performance of Tags}
\begin{figure}[ht]
 \center
 \includegraphics[scale=0.4]{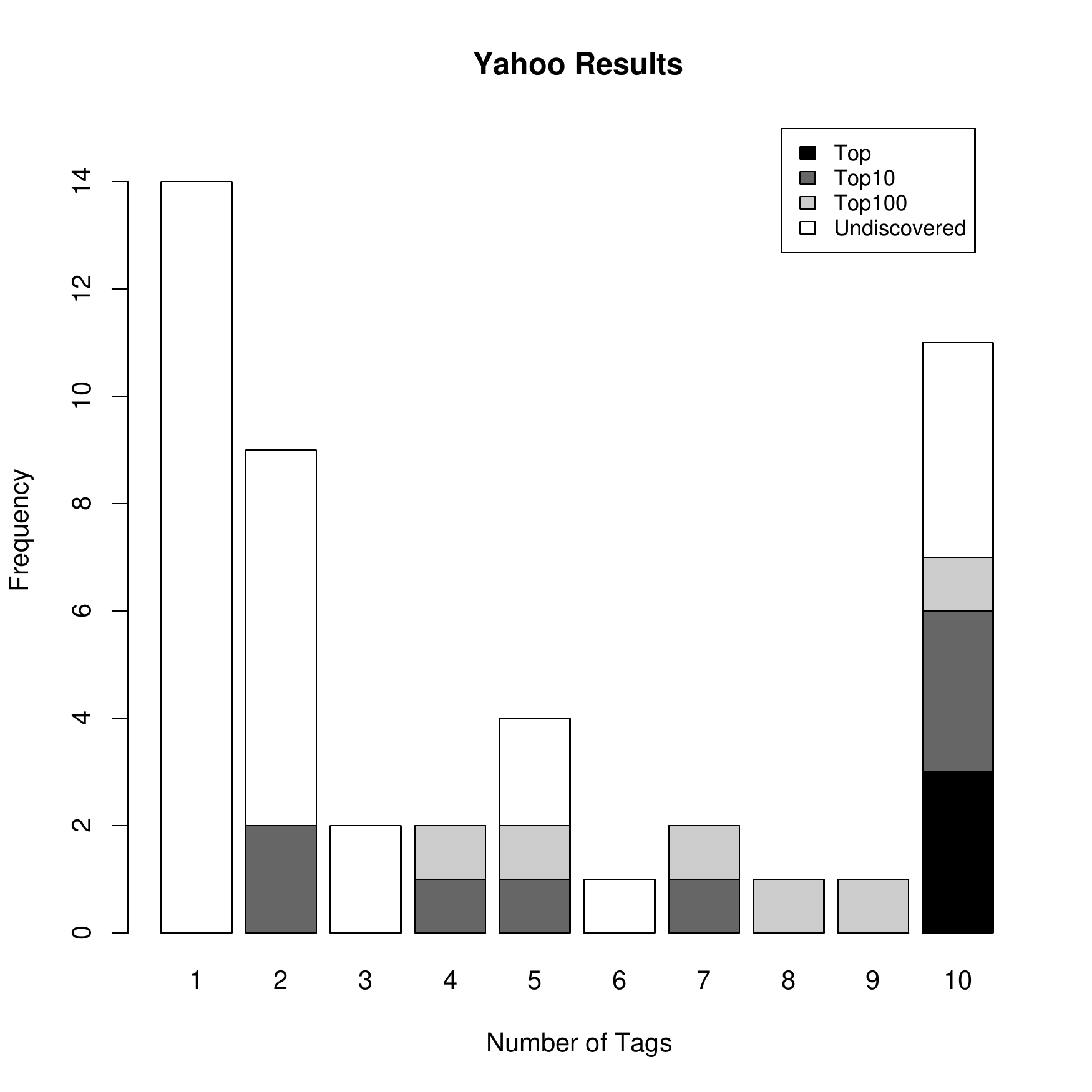}
 \caption{Tags Retrieval Performance by Length in Yahoo!}
 \label{fig:tags_retrieval_yahoo}
\end{figure}

We were able to retrieve tags from $47$ out of our $309$ URIs through the API of the bookmarking website \texttt{delicious.com}.
Not all URIs where annotated with the same number of tags. The API returns at most the top $10$ tags per URI which means 
we need to distinguish between the amount of tags we query against search engines.
The retrieval performance of all obtained tags sorted by their length can be seen in Figure \ref{fig:tags_retrieval_yahoo}.
The length of the entire bar represents the frequency, how many URIs were annotated with that many tags. The shaded
%colored
portions of each bar indicate
the performance. We again distinguish between top ranked, top $10$, $100$ and undiscovered results.
Regardless of how many tags are being used, the retrieval performance in our experiment is poor. Only a few $10$-tag queries actually returned the source
URI as the top result. Figure \ref{fig:tags_retrieval_yahoo} shows results obtained from the Yahoo! API. Since the results are equally bad when querying
tags against Google and MSN Live we do not show those graphs here.
We are aware that the size of our sample set is very limited ($47$ URIs) but we still believe that tags may provide some value for the discovery of missing
pages, especially when titles and LSs are not available.
\subsection{Performance of Neighborhood LSs}
\begin{figure}[ht]
 \center
 \includegraphics[scale=0.4]{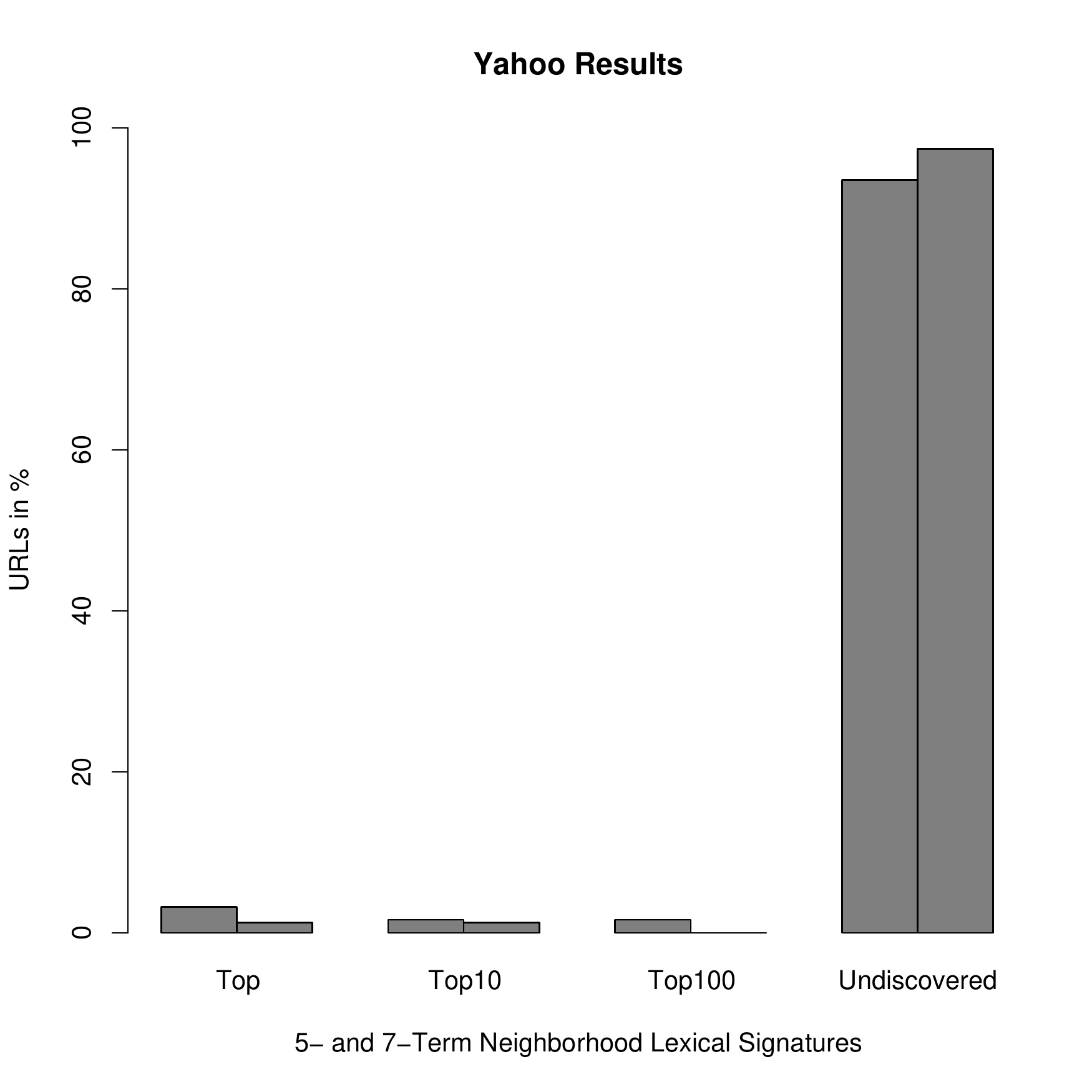}
 \caption{Retrieval Performance of LNLSs in Yahoo!}
 \label{fig:nh_retrieval}
\end{figure}

The results based on LSs generated from the link neighborhood are not impressive. Neither $5$- nor $7$-term LNLSs perform in a satisfying manner.
Slightly above $3\%$ for $5$-term and $1\%$ of all URIs for $7$-term LNLSs are returned top ranked.
As mentioned in Section \ref{subsec:page_nh} we only generated the neighborhood based LSs using the Yahoo! API and also queried the LNLSs only against Yahoo!.
In concurrence with the results seen above, $5$-term LNLSs perform better than LNLSs containing $7$-terms.
Figure \ref{fig:nh_retrieval} shows the relative number of URIs retrieved in the according section of ranks.
\begin{table*}[th]
 \centering
  \begin{tabular}{|c||c|c|c|c||c|c|c|c||c|c|c|c|} \hline
        &\multicolumn{4}{c||}{Google}&\multicolumn{4}{c||}{Yahoo!}&\multicolumn{4}{c|}{MSN Live} \\ 
        &\textbf{Top}&\textbf{Top10}&\textbf{Top100}&\textbf{Undis}&\textbf{Top}&\textbf{Top10}&\textbf{Top100}&\textbf{Undis}&\textbf{Top}&\textbf{Top10}&\textbf{Top100}&\textbf{Undis} \\ \hline \hline
	LS5&50.8&12.6&4.2&32.4&\textbf{67.6}&7.8&2.3&22.3&63.1&8.1&1.6&27.2 \\ \hline
	LS7&57.3&9.1&2.6&31.1&\textbf{66.7}&4.5&1.9&26.9&62.8&5.8&1.6&29.8 \\ \hline
	TI&\textbf{69.3}&8.1&2.9&19.7&63.8&8.1&0.6&27.5&61.5&6.8&1.0&30.7 \\ \hline
	TA&2.1&10.6&12.8&75.5&\textbf{6.4}&17.0&12.8&63.8&0&8.5&10.6&80.9 \\ \hline
  \end{tabular}
  \caption{Relative Number of URIs Retrieved with one Single Method from Google, Yahoo! and MSN Live}
 \label{tab:retrieval_data}
\end{table*}
\begin{table*}[th]
 \centering
  \begin{tabular}{|c||c|c|c|c||c|c|c|c||c|c|c|c|} \hline
        &\multicolumn{4}{c||}{Google}&\multicolumn{4}{c||}{Yahoo!}&\multicolumn{4}{c|}{MSN Live} \\ 
        &\textbf{Top}&\textbf{T10}&\textbf{T100}&\textbf{Undis}&\textbf{Top}&\textbf{T10}&\textbf{T100}&\textbf{Undis}&\textbf{Top}&\textbf{T10}&\textbf{T100}&\textbf{Undis} \\ \hline \hline
	LS5-TI&65.0&15.2&6.1&13.6&\textbf{73.8}&10.0&2.3&14.0&71.5&10.0&1.9&16.5 \\ \hline 
	LS7-TI&70.9&11.7&4.2&13.3&\textbf{75.7}&7.4&1.9&14.9&73.8&9.1&1.9&15.2 \\ \hline 
	TI-LS5&73.5&9.1&3.9&13.6&\textbf{75.7}&9.1&1.3&13.9&73.1&9.1&1.3&16.5 \\ \hline 
	TI-LS7&74.1&9.4&3.2&13.3&\textbf{75.1}&8.7&1.3&14.9&74.1&9.1&1.6&15.2 \\ \hline 
	LS5-TI-LS7&65.4&15.2&6.5&12.9&\textbf{73.8}&10.0&2.6&13.6&72.5&10.4&2.6&14.6 \\ \hline 
	LS7-TI-LS5&71.2&11.7&4.2&12.9&\textbf{76.4}&7.8&2.3&13.6&74.4&9.1&1.9&14.6 \\ \hline 
	TI-LS5-LS7&73.8&9.1&4.2&12.9&\textbf{75.7}&9.1&1.6&13.6&74.1&9.4&1.9&14.6 \\ \hline 
	TI-LS7-LS5&74.4&9.4&3.2&12.9&\textbf{75.7}&9.1&1.6&13.6&74.8&9.1&1.6&14.6 \\ \hline 
	LS5-LS7&52.8&12.9&6.5&27.8&\textbf{68.0}&7.8&2.9&21.4&64.4&8.4&2.6&24.6 \\ \hline 
	LS7-LS5&59.9&9.7&2.6&27.8&\textbf{71.5}&4.9&2.3&21.4&66.7&7.1&1.6&24.6 \\ \hline 
  \end{tabular}
  \caption{Relative Number of URIs Retrieved with Two or More Methods Combined}
 \label{tab:retrieval_improvement}
\end{table*}
\subsection{Combining LSs and Titles}
The observation of well performing LSs and titles leads to the question of how much would we gain if we combined both for the retrieval of the missing page.
To approach this point we took all URIs that remained undiscovered with LSs and analyzed their returned rank with titles (non-quoted only).
Table \ref{tab:retrieval_data} summarizes the results shown in the sections above. It holds the relative numbers of URIs retrieved using one single method.
The first, leftmost column indicates the method. $LS5$ and $LS7$ stand for $5$- and $7$-term LSs, $TI$ for title and $TA$ for tags. Note that for tags we chose to 
display the results for URIs for which we were able to obtain $10$ tags simply because the results were best for those URIs.
The top performing single methods are highlighted in bold figures (one per row).

Table \ref{tab:retrieval_improvement} shows in a similar fashion all combinations of methods that we consider reasonable involving LSs and titles. The reason why
tags are left out here is simple: all URIs returned by tags are also returned by titles and LSs in the according rank section or better. For example, if URI
$A$ is returned at rank five through tags, it is also returned rank five or better with titles and LSs.

The combination of methods displayed in the leftmost column is sensitive to its order, i.e. there is a difference between applying $5$-term LSs first and
$7$-term LSs second and vice versa. The top results of each combination of methods are again highlighted in bold numbers.
Regardless of the combination of methods, the best results are obtained from Yahoo!.
If we consider all combinations of only two methods we find the top performance of $75.7\%$ twice in the Yahoo! results. Once with $LS7-TI$ and once with
$TI-LS5$. The latter combination is preferable for two reasons:
\begin{enumerate}
\item titles are easy to obtain and do not involve a complex computation and acquisition of document frequency values as needed for LSs and
\item this methods returns $9.1\%$ of the URIs in the top $10$ which is $1.7\%$ more than the first combination returns. Even though we do not distinguish
between rank two and rank nine, we still consider URIs returned within the top $10$ as good results.
\end{enumerate}
The combination $LS7-TI-LS5$ accounts for the most top ranked URIs with $76.4\%$. While the $3$-method combinations return good results, there are not significantly
better than for example the two methods mentioned above. The performance delta is not sufficient to justify the expensive generation of LSs without using the
easy to obtain titles first.
The last two rows in Table \ref{tab:retrieval_improvement} show results for combinations of methods based purely on LSs. The results again are good but
it is obvious that a combination with titles (either as the first or second method) provides better results.

%Even though the results from Yahoo! were best, the table also confirms that MSN Live and Google are strong competitors and users would not lose much if
%for some reason tied to using one of them only.
Yahoo! uniformly gave the best results and MSN Live was a close second. Google was third, only managing to outperform MSN Live once ($TI-LS5$) at the top rank.

Since we only have retrieval data for LNLSs from Yahoo!, we isolated all reasonable combinations into Table \ref{tab:nh_data_improvement}.
The first two rows again mirror the results from Figure \ref{fig:nh_retrieval} and the consecutive rows show combinations
of methods. % with LSs and titles.
We can summarize that there is value in combining this method with others but the overall results are not
as impressive as the results shown above. The LNLSs only make sense as a second method in a combination. The reason is very
simple: these kinds of LSs are far too expensive to generate (acquire and download all pages, generate LSs).
% and therefore can only be considered as a secondary option.
This method, similar to tags, can however be applied as a first step in case no copies of the missing page are available in the WI.
\begin{table}
 \centering
  \begin{tabular}{|c||c|c|c|c|} \hline
        &\multicolumn{4}{c|}{Yahoo!} \\ 
        &\textbf{Top}&\textbf{T10}&\textbf{T100}&\textbf{Undis} \\ \hline \hline
	LNLS5&3.2&1.6&1.6&93.5 \\ \hline
	LNLS7&1.3&1.3&0&97.4 \\ \hline
	LS5-LNLS5&68.3&7.8&2.3&21.7 \\ \hline
	LS5-LNLS7&67.6&8.1&2.3&22.0 \\ \hline
	LS7-LNLS5&67.3&4.9&1.9&25.9 \\ \hline
	LS7-LNLS7&66.7&4.9&1.9&26.5 \\ \hline
	TI-LNLS5&64.7&8.1&0.6&26.5 \\ \hline
	TI-LNLS7&64.1&8.7&0.6&26.5 \\ \hline
  \end{tabular}
  \caption{Relative Number of URIs Retrieved from Yahoo! with Methods that Involve LNLSs}
 \label{tab:nh_data_improvement}
\end{table}
\section{Title Analysis and Performance Prediction}
Given that the title of a page seems to be a good method considering its retrieval performance we further investigate the characteristics of our titles.
We analyzed four factors of all obtained titles:
\begin{itemize}
\item title length in number of terms
\item total number of characters in the title and
\item mean number of characters per term
\item number of stop words in the title.
\end{itemize}
Since this series of experiments is also costly on the number of queries, we only ran it against the Yahoo! API.

\begin{figure}[ht]
 \center
 \includegraphics[scale=0.5]{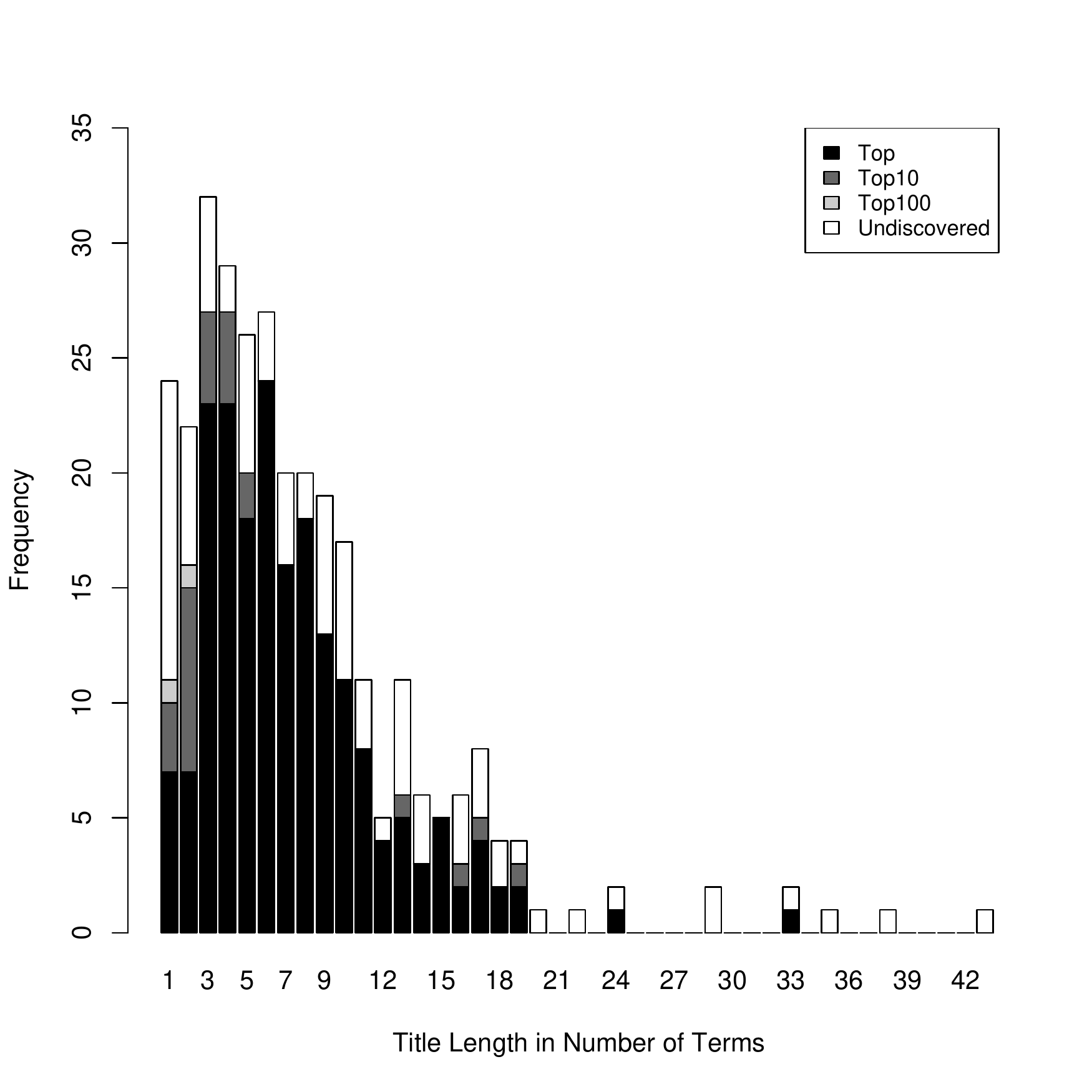}
 \caption{Title Length in Number of Terms vs Rank}
 \label{fig:title_lenvsrank}
\end{figure}

How the title length in number of terms behaves in contrast to the retrieval performance is shown in Figure \ref{fig:title_lenvsrank}.
The setup of the figure is similar to Figure \ref{fig:tags_retrieval_yahoo}. Each occurring title length is represented by its own bar and the 
number of times this title length occurs is shown by the hight of the entire bar. The shaded 
%colored 
parts of the bars indicate how many titles (of the
according length) performed in what retrieval class (the usual, top, top $10$, $100$ and undiscovered).
The titles vary in length between one and $43$ terms. However, there is for example no title with length $21$, hence its bar is of hight null.
Visual observation indicates a title length between three and six occurs frequently and performs fairly well.
Given the data from Figure \ref{fig:title_lenvsrank} we extracted the values for all URIs and are now able to generate a lookup table with the distilled probabilities for
each title length to return URIs in the top, top $10$ and top $100$. We define the probabilities as $P_1$, $P_{10}$ and $P_{100}$.
The lookup table with the probabilities in dependence of the title length (here $TL$) in number of terms is given in Table \ref{tab:title_prob_lookup}.
Using this table, we can predict if a given title is likely to perform well. The predicted probability may have an impact on
what method should be run first. For example, if $P_{1}$ and $P_{10}$ are very low we may want to skip the title query and proceed with LSs right away.
\begin{table}
 \centering
 %{\scriptsize
  \begin{tabular}{|c|c|c|c||c|c|c|c|} \hline
	\textbf{TL}&\textbf{$P_{1}$}&\textbf{$P_{10}$}&\textbf{$P_{op100}$}&\textbf{TL}&\textbf{$P_{1}$}&\textbf{$P_{10}$}&\textbf{$P_{100}$} \\ \hline \hline
	1&0.3&0.4&0.5&2&0.3&0.7&0.7 \\ \hline
	3&0.7&0.8&0.8&4&0.8&0.9&0.9 \\ \hline
	5&0.7&0.8&0.8&6&0.9&0.9&0.9  \\ \hline
	7&0.8&0.8&0.8&8&0.9&0.9&0.9 \\ \hline
	9&0.7&0.7&0.7&10&0.6&0.6&0.6 \\ \hline
	11&0.7&0.7&0.7&12&0.8&0.8&0.8 \\ \hline
	13&0.5&0.5&0.5&14&0.5&0.5&0.5 \\ \hline
	15&1.0&1.0&1.0&16&0.3&0.5&0.5 \\ \hline
	17&0.5&0.6&0.6&18&0.5&0.5&0.5 \\ \hline
	19&0.5&0.8&0.8&24&0.5&0.5&0.5 \\ \hline
	33&0.5&0.5&0.5&&&& \\ \hline
  \end{tabular}
%}
  \caption{Lookup Table for Performance Probability of Titles Depending on Their Length}
 \label{tab:title_prob_lookup}
\end{table}

\begin{figure}[ht]
 \center
 \includegraphics[scale=0.5]{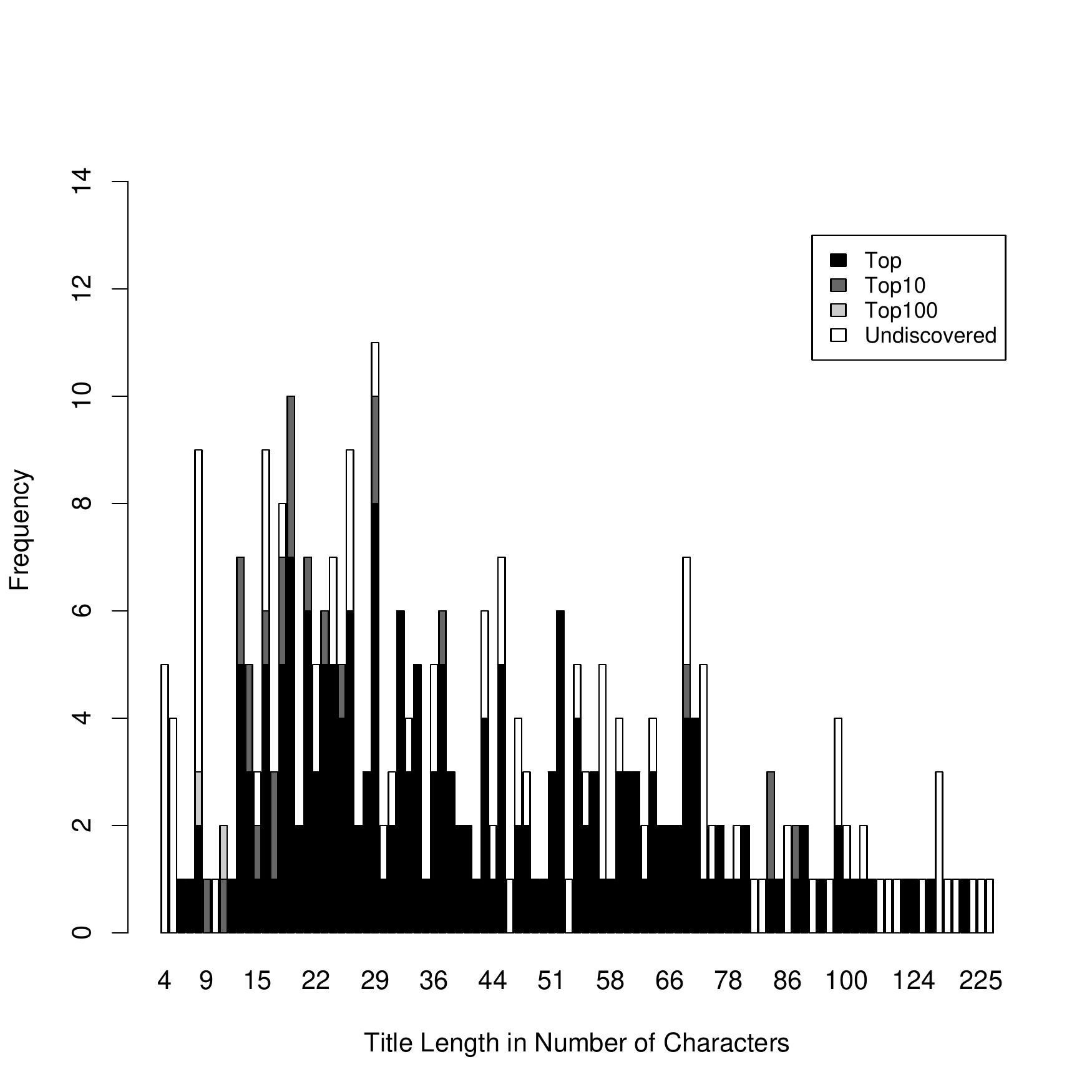}
 \caption{Title Length in Number of Characters vs Rank}
 \label{fig:title_clenvsrank}
\end{figure}

The contrast of total title length in number of characters and rank is shown in Figure \ref{fig:title_clenvsrank}.
While the title length varies greatly between $4$ and $294$ characters we only see $15$ URIs with a title length greater or equal to
$100$ and only three URIs with more than $200$ characters in their title.
Figure \ref{fig:title_clenvsrank} does not reveal an obvious pattern between number of characters and rank returned for a title but very short titles
(less than $10$ characters) do not seem to perform well. A title length between $10$ and $70$ characters is most common and the ranks seem to be better in
the range of $10$ to $45$ characters total.

\begin{figure}[ht]
 \center
 \includegraphics[scale=0.45]{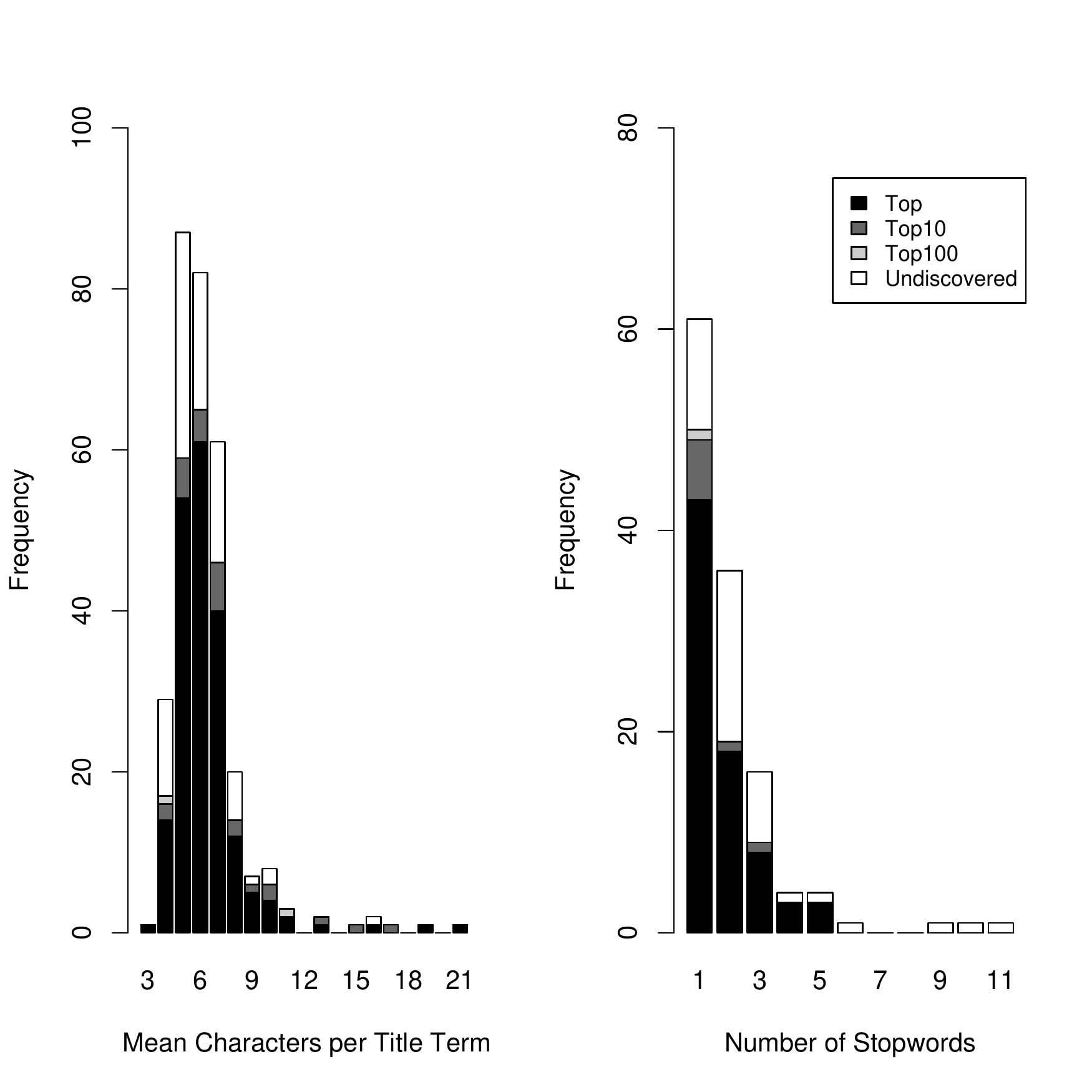}
 \caption{Mean Number of Characters per Title Term and Number of Stop Words vs Rank}
 \label{fig:title_sw_meanlenvsrank}
\end{figure}

Figure \ref{fig:title_sw_meanlenvsrank} depicts on the left the mean number of characters per title term and their retrieval performance.
It seems that terms with an average of $5$, $6$ or $7$ characters seem to be most suitable for well performing query terms.
On the bottom right end of the barplot we can see two titles that have a mean character length per term of $19$ and $21$. Since such long words are
rather rare they perform very well. 

The observation of stop word frequency in the titles and their performance is not surprising. As shown on the right in Figure \ref{fig:title_sw_meanlenvsrank}
titles with more than a couple of stop words seem to harm the performance. The intuition is that search engines filter stop words from the query
(keep in mind, these are non-quoted titles) and therefore it makes sense that for example the title with $11$ stop words does not return its URI within
the top $100$ ranks.

\begin{figure}[ht]
 \center
 \includegraphics[scale=0.45]{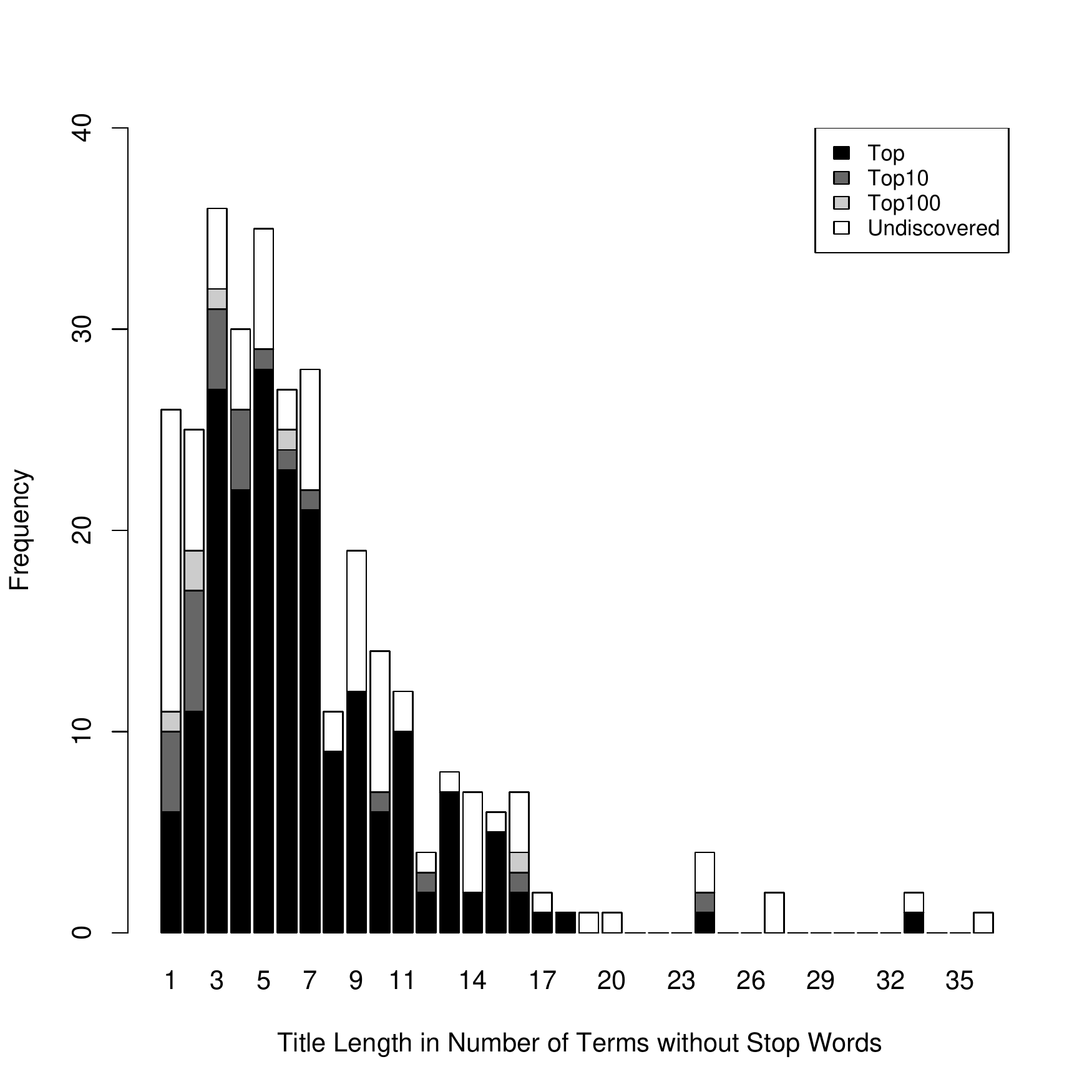}
 \caption{Title Length in Number of Non-Stop Words vs Rank}
 \label{fig:title_nosw_lenvsrank}
\end{figure}

For completeness we removed all stopwords from the titles and analyzed their retrieval performance in dependence of the new title length.
The results are shown in Figure \ref{fig:title_nosw_lenvsrank}. As expected we see more titles with fewer terms performing slightly better
than the original titles.
This result indicates that the performance of the method using the web page's titles can still be improved. Further analysis of the 
best combination of methods with titles without stop words remains for future work.
\section{Future Work} \label{sec:futwork}
Our main aspect of future work is the implementation of the system described in the flow diagram of Figure \ref{fig:flow_diagram}.
The system will operate as a browser plugin and will trigger once the user encounters a $404$ ``Page Not Found'' error.
It will provide all of the introduced methods to rediscover the missing page and since the discovery process happens in an automated 
fashion, the system can provide the user with the results in real-time while she is browsing the Internet.

We have shown in \cite{klein:ls} that LSs evolve over time and consequently lose some of its retrieval strength. Here we are arguing that
titles of web pages are a powerful alternative to LSs. The next logical step is to investigate the evolution and possible decay of titles
over time. Our intuition is that titles do not decay quite as quickly as LSs do since the actual content of a web page (a headline, sentence
or paragraph) presumably changes more frequently than its general topic which is what the title is supposed to represent.

Our set of obtained tags is limited. It remains for future work to investigate the retrieval performance of tags in a large scale experiment.
It also would be interesting to see what the term overlap between tags, titles and LSs is since all three methods are generated on different
grounds.

Our method to generate LNLSs may not be optimal. We chose to use inlink pages only 
%since they supposedly represent the content of the centroid page better than outlink pages.
%We also created a bucket of all neighborhood terms per URI and generated the LNLSs based on this bucket.
and created a bucket of all neighborhood terms per URI. The LNLSs are based on this bucket.
It remains to be seen whether outlink pages actually can contribute to the retrieval performance and other methods than the bucket
of terms are preferable.
It is possible that our neighborhood is too big since it includes the entire neighboring page. A page that links to many pages (hub) %as well as ours
may have a diffuse ``aboutness''. Hence we are going to restrict the content gained from the neighborhood to the link anchor text of the inlink
pages.
\section{Conclusions} \label{sec:concl}
In this paper we evaluate the retrieval performance of four methods to discover missing web pages. 
We generate a dataset of URIs by randomly sampling URIs from \texttt{dmoz.org} and assume these pages to be missing.
We generate LSs from copies of the pages, parse the pages' titles, obtain tags of the URIs from the bookmarking website
\texttt{delicious.com} and generate LSs based on link neighborhood.
We use the three major search engines Google, Yahoo! and MSN Live to acquire mandatory document frequency data for the generation of the LSs.
We further query all three search engines for all our methods and combine methods to improve the retrieval performance.
We are able to recommend a setup of methods and see one search engine performing best in most of our experiments. 

It has been shown in related work that LSs can perform well for retrieving web pages. Our results confirm these findings, for example 
more than two-thirds of our URIs have been returned as the top result when querying $5$- and $7$-term LSs against the Yahoo! search engine API.
They also lead us to the claim that titles of web pages are a strong alternative to LSs. Almost $70\%$ of the URIs have been returned as
the top result from the Google search engine API when queried with the (non-quoted) title.
However, our results show that a combination of methods performs best. Querying the title first and then using the $5$-term LSs 
for all remaining undiscovered URIs against Yahoo! provided the overall best result with $75.7\%$ of top ranked URIs and
another $9.1\%$ in the top $10$ ranks.
The combination $7$-term LS, title, $5$-term LS returned $76.4\%$ of the URIs in the top ranks but since LSs are more expensive
to generate than titles, we recommend the former combination of methods. A good strategy, based on our results, is to query the title
first and if the results are insufficient generate and query LSs second. Yahoo! returned the best results for all combination of methods and thus seems
to be the best choice even though Google returned better results when querying the title only.
\section{Acknowledgement} \label{sec:concl}
This work is supported in part by the Library of Congress.
%\bibliographystyle{abbrv}
%\bibliography{jcdl2010}  % sigproc.bib is the name of the Bibliography in this case
% ACM needs 'a single self-contained file'!
%

%
% That's all folks!
\end{document}